\documentclass[a4paper]{article}

\usepackage{cite}
\usepackage{amsmath,amssymb,amsfonts}
\usepackage{algorithmic}
\usepackage{graphicx}
\usepackage{textcomp}
\usepackage{booktabs} 
\usepackage{hyperref}
\usepackage{cleveref}
\usepackage{multirow}
\usepackage{color}
\usepackage{diagbox}
\usepackage{xcolor,colortbl}
\usepackage{enumitem}
\usepackage{balance}
\newcommand{\eg}{e.\,g.,\,}

\newcommand{\cf}{{cf.\,}}

\usepackage[activate]{microtype}
\newcommand{\opensmile}{\textsc{openSMILE\,}}
\newcommand{\lld}{\textit{low-level descriptors\,}}
\newcommand{\mfcc}{\textit{Mel-frequency cepstral coefficients\,}}
\newcommand{\cmp}{\textsc{ComParE\,}}
\newcommand{\ds}{\textsc{Deep Spectrum\,}}
\newcommand{\dss}{\textsc{DS\,}}
\newcommand{\egm}{\textsc{eGeMAPS\,}}
\newcommand{\score}{\textsc{Score!\,}}

\usepackage{soul}

\definecolor{greenish}{rgb}{0.0,0.7,0.0}

\usepackage{INTERSPEECH2020}

\title{Predicting Sex and Stroke Success --\\ Computer-aided Player Grunt Analysis in Tennis Matches}
\name{Lukas Stappen$^1$, Manuel Milling$^1$, Valentin Münst$^1$, Korakot Hoffmann$^1$, Bj{\"o}rn W.\ Schuller$^{1,2}$}

\address{
  $^1$Chair of Embedded Intelligence for Health Care and Wellbeing, University of Augsburg, Germany.\\
  $^2$GLAM -- the Group on Language, Audio, \& Music, Imperial College London, UK}
\email{stappen@ieee.org}

\begin{document}

\maketitle
\begin{abstract}
\vspace{-0.1em}
Professional athletes increasingly use automated analysis of meta- and signal data to improve their training and game performance. As in other related human-to-human research fields, signal data, in particular, contain important performance- and mood-specific indicators for automated analysis. In this paper, we introduce the novel data set \score to investigate the performance of several features and machine learning paradigms in the prediction of the sex and immediate stroke success in tennis matches, based only on vocal expression through players' grunts. The data was gathered from YouTube, labelled under the exact same definition, and the audio processed for modelling. We extract several widely used basic, expert-knowledge, and deep acoustic features of the audio samples and evaluate their effectiveness in combination with various machine learning approaches. In a binary setting, the best system, using spectrograms and a Convolutional Recurrent Neural Network, achieves an unweighted average recall (UAR) of $84.0\,\%$ for the player sex prediction task, and $60.3\,\%$ predicting stroke success, based only on acoustic cues in players' grunts of both sexes. Further, we achieve a UAR of $58.3\,\%$, and $61.3\,\%$, when the models are exclusively trained on female or male grunts, respectively. 
\end{abstract} 

%In particular, Long Short-term Memory Recurrent (LSTM-RNN) and Convolutional Recurrent Neural Network (CRNN)  are trained from scratch on the basic, multidimensional features, whilst the compressed feature representations (expert-knowledge and deep features) are utilised by a linear Support Vector Machine (SVM). 

\vspace{-0.1em}
\section{Introduction\label{sec:Introduction}}
\vspace{-0.1em}
Non-verbal vocalisation -- namely screams, moans, and grunts -- is omnipresent in human communication. However, the understanding and differentiation of communicative functions is underrepresented in research. Despite the widespread use of computational analyses in sports, so far only little attention has been paid to vocalisations. One of the most plausible reasons for this is the complex nature of vocalisations and the abundance of noise in sports. The success of audio analysis in affective computing, however, suggests that valuable information about the emotional state of individuals can be extracted from the voice \cite{baird2021,schuller2017interspeech, schuller2018interspeech}.%,schuller2017interspeech 

The concept of complex computational analysis in the world of sports has gained massive popularity over the past decade~\cite{link2009sport}. A prominent example is the successful Oakland team in Major League Baseball, which relies largely on statistical data analysis, rather than subjective scouting reports, when buying new players~\cite{lewis2003moneyball}. %Nowadays, the scope of computational assistance covers opponent analysis for game preparation~\cite{hui2010computer}, as well as evaluation systems for faults in sports techniques%sensor data evaluation for regeneration, and camera-based movement analysis
%~\cite{}. %, brzostowski2013adaptive
For instance, in tennis, a sport enjoying worldwide attention, player tracking systems based on computer vision have previously been developed to calculate statistics and to analyse performance~\cite{nieto2013automatic}. Further, approaches based on metadata have been used to predict the probability of players scoring with their serve~\cite{sipko2015machine}.  

In this work, we study the computer-aided, acoustic classification of grunts -- a non-verbal screaming vocalisation made by tennis players across sexes when striking the ball.
The early work of~\cite{sinnett2010preliminary} demonstrated experimentally that grunting makes it more difficult for opponents and observers to determine the probable trajectory of the ball, so the technique can be used for distraction and concealment.~\cite{Connell2014} found that the screams of the players significantly increased the velocity and power of a shot, while~\cite{raine2017tennis} investigated typical frequencies of the shouts during strokes and serves, in relation to the sex, age, and height of the players. In particular, differences in acoustic characteristics were found for sex and stroke type showing that certain acoustic cues in tennis grunts correlate with the vocaliser's sex and hit type and indicate the outcome of a match. Recent research suggests that tennis grunts contain audio information which can be systematically exploited~\cite{muller2019sound}. 
However, all these approaches lack an automated approach to identifying and leveraging these acoustic features for prediction. 
%, which is in line with a study by~\cite{raine2017tennis} 

The hypothesis of our work is that acoustic tennis grunts contain generalisable patterns, so that we are able to learn acoustic cues automatically using advanced audio features and machine learning approaches. For this reason, we selected, collected, and annotated a novel data set of real-world tennis matches from YouTube. In addition, we extracted a wide range of basic audio features (\eg \lld (LLDs), \mfcc (MFCCs)), expert-knowledge features (the \textit{extended Geneva Minimalistic Acoustic Parameter Set} (\egm) and \cmp), and state-of-the-art deep representations (\ds (DS)) to train suitable neural network architectures and Support Vector Machines (SVMs). These features have shown their efficiency on several non-verbal vocalisation tasks \eg predicting baby sounds~\cite{schuller2020interspeech}, crying~\cite{schuller2018interspeech}, and snoring~\cite{schuller2017interspeech}. Furthermore, \dss recently demonstrated promising performances in multiple audio-based tasks~\cite{schuller2020interspeech, stappen2021}. % Amiriparian19-AYP, Amiriparian18-BND

Equipped with these comprehensive tools, we studied the automatic prediction of a) the \textit{sex} of the grunt vocaliser, and b) whether a player \textit{immediately scores a point}, solely based on their grunt during the stroke. To the authors' knowledge, this is the first time such a study has been conducted on a tennis grunting corpus. On task a) we achieved an Unweighted Average Recall (UAR) of $84.0\,\%$ and on b) a UAR of $60.3\,\%$ (both sexes) on average over a 5-fold player-independent cross partition (chance level $50\,\%$).
Our results indicate that some immediate information about the physiological and psychological state of the players is communicated through tennis grunts, which shows further potential for deeper investigations.

%The remainder of this paper is organised as follows. We  provide a brief overview of relevant related work covering the topics of non-verbal vocalisations and sports analyses in~\Cref{sec:RelatedWork}. In~\Cref{sec:DataSet}, our novel tennis grunt data set is introduced, including details about the data collection and annotation process. Next, we explain our extracted feature sets and machine learning approaches in~\Cref{sec:Modelling}, based on which we obtained the results presented in ~\Cref{sec:Results}. We conclude this work with a summary of our findings and future work in~\Cref{sec:Conclusion}.

\vspace{-0.1em}
\section{Related Work\label{sec:RelatedWork}}
\vspace{-0.1em} %lohr2012age
In recent years, a variety of innovative, computer-based methods, \eg the Internet of Things and Big Data analytics, found new applications in sports sciences, leading to the emergence of sports informatics~\cite{rein2016big, ray2015internet}.
%~\cite{link2009sport}.
Improvements in hardware, such as wearables, enable scientists to gather large quantities of data from realistic sport environments~\cite{sykora2015advances}. Combined with machine learning techniques, this trend has led to new possibilities in supporting and advancing theory and practice in sports~\cite{cao2012sports}. For example, shot~\cite{draschkowitz2015using} and match~\cite{schumaker2010predictive} outcomes are predicted in table tennis and basketball, before or during matches using video and historical data. Moreover, variables with strong predictive power~\cite{sipko2015machine} can be automatically identified. %bunker2019machine   and process

Specifically for tennis, the winner of a match is predicted with an accuracy of up to 80\,\%~\cite{somboonphokkaphan2008tennis}, based on data from past Grand Slam tournaments, utilising a multilayer perceptron approach. In~\cite{panjan2010prediction}, morphological measures and motor tests were able to estimate a tennis player's performance, showing positive results with linear regression.
Further in-depth studies analysed tennis shots, recommending target locations for optimal serves~\cite{whiteside2017spatial} and predicting the serve type of individual players in a given context, based on Hawk-Eye data~\cite{wei2015predicting}. Further research dealt with the emotions of tennis players, which are expressed through the posture and gestures of the player~\cite{calvo2015expressing, nardelli2015recognizing}. These studies indicate that players' faces are significant indicators of their emotions after the outcome of a point. Recently,~\cite{kovalchik2018going} found that several emotional states influence the player's chances of scoring the next point. 
% And, you do not relate to the Interspeech Challenge in 2010 where we did gender as task - this one summarises it: https://www.sciencedirect.com/science/article/abs/pii/S0885230812000162

Acoustics are well known for containing paralinguistic information and, therefore, being a useful indicator of psychological states, emotions~\cite{schuller2017interspeech}, and gender~\cite{Connell2014, Schuller13PSL}. In a paralinguistic challenge, participants showed the potential of predicting an individual's sex based on speech utterances. A detailed description of the feature sets and models can be found in~\cite{Schuller13PSL}. Humans subconsciously change their voice pitch, \eg when meeting people for the first time~\cite{cheng2016listen} or talking to people of a higher social status~\cite{leongomez2017perceived}, to feel more dominant and powerful. In psychology, these placebo expectations are expected to be a self-fulfilling prophecy of objective reality~\cite{crum2015self}. However, non-verbal vocalisations have received little attention in sports science research, despite having previously proven their suitability for various complex tasks.~\cite{Raine19-HRC} found that listeners deduce certain physical properties from non-verbal vocalisation types, and roars are well suited for predicting the height and upper-body strength of a human individual. In the context of sports, statistical methods have been used to find predictors between tennis players' grunts and their psychological properties, as well as linking grunts to nonhuman mammal calls~\cite{raine2017tennis}. The authors concluded that tennis grunts communicate information about the player's sex, forehand and backhand strokes, and the outcome of a match.
%~\cite{stel2012lowering}

%The authors concluded that tennis grunts communicate information about the players as well as the contest, given that they were able to find a correlation between the grunts and the outcome of the match. Further analysis indicated differences in grunts depending on the shot type. The grunts recorded during serves had a higher fundamental frequency than the ones recorded during forehand and backhand shots.
%Convolutional Neural Networks (CNNs) have previously shown good performance on acoustic data sets~\cite{hershey2017cnn, valenti2016dcase}.

\vspace{-0.3em}
\section{\score -- A Tennis Grunt Dataset\label{sec:DataSet}}

% Since this is the first? data set targeting to automatically predict scores on sounds we had to create a suitable data set. 

\subsection{Data Collection and Selection}
\vspace{-0.1em}
We utilised the video platform, YouTube, to collect audio-visual content from 50 real tennis matches to create the novel audio data set \score. The content was identified by searching manually, using keywords related to tennis and grunting and contain mostly full tennis matches equally of women and men players. Since the recordings are from a wide variety of YouTube channels, the type and number of microphones used is unknown. However, we only considered professional matches, in which the audio signal was recorded with additional microphones placed close to the boundaries of the tennis court. The normalised recordings are stored with a 44.1\,kHz sampling frequency and 16\,bits amplitude resolution.

All videos were carefully sifted by hand and processed further if both of the players have grunted in the match, while the background noise occurring simultaneously with the grunts is as low as possible. The latter is especially important, since a low level of audio noise makes it easier for future models to distinguish expressive audio characteristics from meaningless ones. Typical noise patterns we observed are net shots, the audience, referee calls, and the ball hitting the racket. 
We aimed to maintain a high diversity of players, as well as an equal number of audio samples for each player for annotation, to increase the speaker- (or in this case grunting-) invariance. %~\cite{senior2014improving, luo2018speaker}
One player per match and only one video per player was considered for annotation. This forces the models to learn generalisable acoustic features and not individual player or recording-related patterns, such as, a recording device type or certain distances from the court to the additional microphone(s). To ensure a minimum of 30 grunts from each player, we limit our selection to video clips longer than 20 minutes.
Furthermore, we found that the number of grunts per match is highly dependent on the player, but generally both the number of grunts and the volume increase over the course of a match, which is in line with~\cite{raine2017tennis}'s observation regarding the fundamental frequency (F0). To avoid a skew towards higher intensity parts of a match, we sampled the grunts uniformly across the duration of the video. In total, we identified 20 different matches suitable to annotate 600 grunts. Our data set is almost twice as comprehensive as that of~\cite{raine2017tennis}, used for statistical analysis.

\begin{figure*}[tp]
    \centering
    \includegraphics[width=0.95\textwidth]{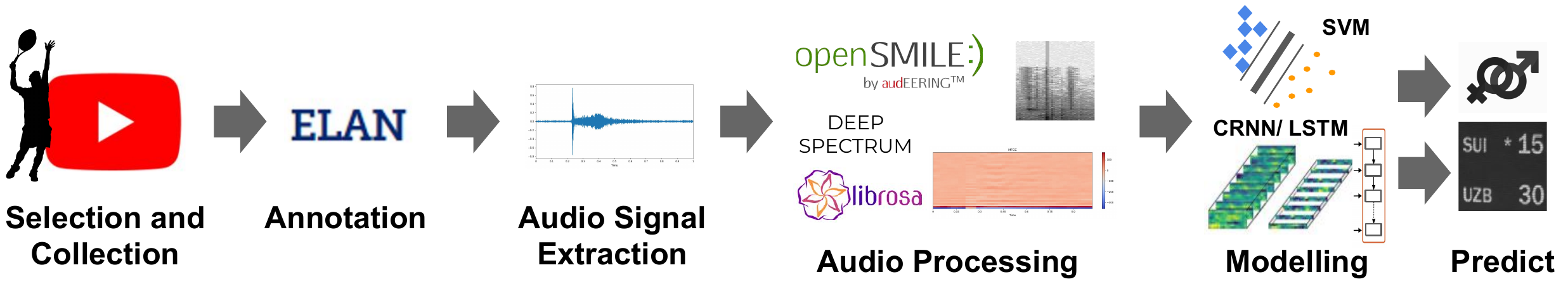} \vspace{-.8em}
    \caption{An overview of our database creation process, preprocessing, and modelling. First, the video data is collected, annotated, and the audio signal separated. From the audio recordings, several features, such as spectrograms and MFCC, are extracted by established frameworks. Afterwards, using SVMs and neural network predictions for the sex and the stroke success detection tasks are obtained. For a detailed account on the pipeline, we refer to~\Cref{sec:Modelling}.}
    \label{fig:DataProcessingSequence}
    \vspace{-1.8em}
\end{figure*}
\vspace{-0.1em}

\vspace{-0.3em}
\subsection{Data Annotation and Partitioning}
\vspace{-0.2em}
After the collection, the data was annotated by two experts and counter-checked by a third person, considering the following criteria. We annotated whether the player who played the stroke and uttered a grunt scored or not. More precisely, the \textit{score} label is defined as: i) \textit{score:} the ball is played inbound and the opponent cannot return the ball or is forced to fault; \textit{not scored:} the ball can be returned by the opponent in accordance with the rules or the ball does not hit the court of the opposing team (e.\,g., net, out of bound, fault when serving). The labels are evenly distributed for each player, so half of the 30 grunts per game led to a point and the other half did not. The second label \textit{sex} indicates if the grunts were uttered by a male or a female player.  
The data was labelled with the annotation software ELAN \footnote{ELAN (Version 5.2) https://tla.mpi.nl/tools/tla-tools/elan/}. The preparatory analysis of the annotation showed that the optimal length of the annotation is around 1000 milliseconds. A longer duration would capture unwanted background noise, a shorter one would not cover the full length of some grunts. After the annotation process, the data was partitioned to equally sized, player-independent 5 folds for cross validation. This approach allows more data to be utilised for the training of the models, while increasing the robustness of the results~\cite{wong2015performance}. 
%cawley2010overm}
\vspace{-0.4em}
\section{Modelling and Experimental Settings\label{sec:Modelling}}
\vspace{-0.1em}
% Most of the process was done manually to achieve a high data quality under consistent definition. 

An overview of our pipeline is given in~\Cref{fig:DataProcessingSequence}. First, basic features are extracted from the audio signals, \eg spectrograms, and low-level descriptors (cf.~\Cref{sec:features}). This step is followed by advancing the features for our tasks. In the final step, machine learning models (\eg deep neural networks and SVMs) receive the features and predict the previously defined labels of \textit{sex} or \textit{score} (cf.~\Cref{sec:models}). All annotations, features, models, and codes are publicly available\footnote{www.github.com/lstappen/score}.

\vspace{-0.5em}
\subsection{Features\label{sec:features}}
\vspace{-0.2em}
In order to classify events from audio signals, feature sets are needed whose extraction is based on decades of research in the field of sound processing. Since automatic tennis grunt analysis is unexplored from the point of view of machine learning, we decided to run extensive experiments on different types of feature sets. The most common basic features originated in audio processing engineering. Over the years, these low-level features have been further developed for specific tasks, such as emotion recognition\cite{Eyben15-TGM,schuller2018interspeech}. From these, expert knowledge features can be calculated. Advancing the features demonstrated higher suitability for many tasks, especially on smaller data sets~\cite{schuller2020interspeech,schuller2018interspeech, schuller2017interspeech}. However, recent achievements in representation learning, which aims to learn features directly based on low-level features well suited to the recognition task, have shown that robust feature sets of audio signals in the context of a particular task can also be learnt automatically~\cite{Amiriparian17-SSC, stappen2020}. %valstar2016avec 

\vspace{-0.5em}
\subsubsection{Basic audio features\label{sec:basicfeatures}}
\vspace{-0.25em}
We extracted LLDs using the toolkit \opensmile~\cite{Eyben13-RDI}. For this purpose, the audio files were downsampled and converted to 16\,kHz (mono, 16\,bit) for a 10\,ms frame-level extraction. The resulting 130-dimensional \textsc{ComParE}-\textit{LLDs} consist of 55 spectral, 4 energy-related, and 6 voice-related features and their first derivatives. In addition, we extracted 40 \textit{MFCCs} based on the originally sampled (44.1\,kHz) audio files using the Python package, Librosa. Finally, applying a periodic Hanning window with a width of 16\,ms and an overlap of 8\,ms we extracted 227x227-dimensional \textit{spectograms} from audio. %~\cite{ittichaichareon2012speech}

\vspace{-0.5em}
\subsubsection{Expert-knowledge features for sound classification}
\vspace{-0.25em}
A prevailing `traditional' approach to sound sensing consists of combining LLDs with a set of statistical measures calculated over a fixed duration sliding window. In a series of audio challenges~\cite{schuller2020interspeech,schuller2018interspeech,schuller2017interspeech}, \textit{\cmp} functionals demonstrated the suitability to extract meaningful features for a wide range of tasks. 
This set consists of 6\,373 static features derived from the calculation of statistics over \textsc{ComParE}-LLD contours. %~\cite{Schuller13-TI2}
Further, we chose the \egm because it is widely used, with a proven track record of predicting physical arousal and valence~\cite{Eyben15-RSA}. %\cite{Eyben15-TGM}
%A detailed description of the feature set can be found in~\cite{Eyben15-RSA}.
It comprises 88 measures covering the LLDs as dimensional spectral, cepstral, and prosodic measures, \eg jitter, pitch, or loudness, including eight functional measures (percentiles, mean, standard deviation, slope, time peaks) from the latter two. These features were extracted with a smoothed symmetrical moving average filter of 3 frames length.

\vspace{-0.5em}
\subsubsection{Deep representation}
\vspace{-0.25em}
Recently, deep representation features performed well on several audio-based event detection as well as emotion recognition tasks~\cite{schuller2020interspeech, Amiriparian17-SAU}. We used the \ds toolkit\footnote{\url{https://github.com/DeepSpectrum/DeepSpectrum}}~\cite{Amiriparian17-SSC} to compute \textit{deep representations} of the audio samples based on extracted colormap-converted spectrograms (\cf \Cref{sec:basicfeatures}). The model for the deep feature extraction is an AlexNet
%~\cite{Simonyan2014} 
pre-trained on ImageNet dataset. As features, the activations of the second fully connected layer are extracted. 
\vspace{-0.4em}

\subsection{Machine learning models\label{sec:models}}
\vspace{-0.1em}
\subsubsection{Support Vector Machine}
\vspace{-0.25em}
Due to their scalability on high dimensional data, SVMs, in combination with expert-knowledge features~\cite{schuller2020interspeech} and deep representations~\cite{ Amiriparian17-SSC}, showed results comparable to most state-of-the-art approaches, especially on smaller and medium-sized datasets. The model is trained using 1000 iterations and a complexity value $C$ between $10^{-5}$ to $10^{1}$.

\vspace{-0.5em}
\subsubsection{Long Short-term Memory Network}
\vspace{-0.25em}
Many basic features are derived from a signal sampling point of view but do not reflect any temporal changes of the signal. In order to learn these, we utilised two stacked layers of Long Short-term Memory Recurrent Neural Network (LSTM-RNN).%~\cite{hochreiter1997long}. 

%\subsubsection{CNN}
%\label{ssec:cnn}

%The four convolutional blocks have a $2\times2$ convolutional kernel with 16, 32, 64 and 128 filters, a max-pooling operation with a size of 2 for the time dimension, as well as, batch normalisation and a dropout of 20\%~\cite{Srivastava2014}. For the final representation, we apply a global average-pooling operation, summarising the strongest signal peaks. The small size of the sliding windows means that the features extracted mostly contain short-term temporal context.
\vspace{-0.5em}
\subsubsection{Convolutional Recurrent Neural Network\label{ssec:crnn}}
\vspace{-0.25em}
The inherent nature of the Convolutional Neural Network (CNN) limits the ability to learn long-term temporal context. To overcome this, a combination of CNNs and RNNs has proven to be suitable for emotion and audio event detection tasks~\cite{rizos2019modelling, tzirakis2019real}. First, the one-dimensional, convolutional layers learn filters shifted over a single spatial dimension, which enables them to extract high-level features that are shift-invariant.%~\cite{norouzi2009stacks,abdel2014convolutional}
The CNN blocks are followed by two bidirectional LSTM-RNNs, where the latest hidden representation from the second layer is fed into a fully-connected classification layer.
Our CRNN possesses three convolutional blocks, each consisting of a one-dimensional convolution layer with ReLU activation, a max pooling layer, and a dropout layer applying 0.5 dropout, and two doubly stacked LSTM-RNN layers. We applied convolutional filter sizes of 10, 20, 40 (kernel: 6, 8, 10) for both, MFCCs and spectrograms, as well as filter sizes of 30, 30, 40 (kernel: 10, 8, 10) for LLDs. 

\vspace{-0.4em}
\section{Results and Discussion\label{sec:Results}}
\vspace{-0.1em}
Spectrograms seem to be very well suited for the task of predicting the \textbf{sex of a grunt} (\cf \Cref{tab:results_sex}). In this regard, they yielded the best results of all features, both in combination with the CRNN ($84.0\,\%$ UAR) and the SVM ($82.3\,\%$ UAR) architecture. The \cmp functionals follow closely behind, with a UAR of $79.5\,\%$ and exceed the other precalculated feature sets, \dss and \egm. It is also noteworthy that a UAR of $> 71$\,\% could be achieved on the basis of a single extracted basic feature vector (middle). This is in line with~\cite{raine2017tennis}, which assumed the predictive power of the peak F0 (LLD) in the prediction of the player's sex. However, the flattened features that cover the entire audio sequence are even more predictive, but lead to extremely large dimensions of the input vectors ($13\,000$ (LLD), $1\,760$ (MFCC), $51\,529$ (spectrogram)) in an SVM setting. Although SVMs validated their strength with high-dimensional vectors and comparatively modest amounts of data, altogether the neural architectures (with the exception of the LLDs) performed better, 
%BS: !!! Something is missing here? 
were more resource efficient and proved the potential of advanced modelling for this particular task. Further, the best feature model combination slightly overemphasised female grunts, predicting 12.5\,\% of male grunts wrongly as female while only 3.5\,\% of women were falsely predicted as men. 

% MFCC values are not very robust in the presence of additive noise, and so it is common to normalise their values in speech recognition systems to lessen the influence of noise. modifications to the basic MFCC to improve robustness, such as by raising the
% raw audio signal did not work

\begin{table}[t!]
 	\caption{
 	    Results over player-independent 5-fold cross validation on the prediction of sex. 
 	    Displayed hyperparameters:
      	$C$: Complexity parameter of the SVM, optimised for all from $10^{-5}$ to $10^{1}$;
      	$lr$: Learning rate of the neural networks. 
      	Three types of aggregation:
      	(middle) token of the sequence;
      	(mean) all tokens averaged along the time axis; 
      	(flat) concatenated token along the time axis. 
		We report the averaged ($\O$) \textbf{UAR}: Unweighted Average Recall and the standard deviation ($\pm$).}
	\vspace{-0.8em}
	\label{tab:results_sex}
\centering
  \resizebox{\columnwidth}{!}{
    \begin{tabular}{l|lrr} %||rr
    \toprule
    %                &           & \multicolumn{2}{c}{combined} \\
    \textbf{} & $C$  &  \O [\%] & +-  [\%] \\   \hline  %& \max [\%] & \pm [\%]
     \multicolumn{4}{c}{\bf  Features (aggregation) + SVM} \\ \hline  
    \textbf{\textsc{LLDs (mean)}} & $10^{-2}\ast$	 &  72.3  & 14.8 \\ %72.33  & 14.78 & 78.50	& 13.87 
    \textbf{\textsc{LLDs (middle)}} & $10^{-1}\ast$	 &  71.5  & 8.2 \\ %&  71.50  & 8.17 & 72.67 &	7.63
    \textbf{\textsc{LLDs (flat)}} & $10^{-3}\ast$	 &  80.5 & 12.3 \\ %&  80.50 & 12.33 & 72.67 &	7.63
    \textbf{\textsc{MFCCs (mean)}} & $10^{1}\ast$	 &  69.5  & 12.4 \\ %&  69.50  & 12.38& 74.17	& 9.97
    \textbf{\textsc{MFCCs (middle)} } & $10^{1}\ast$ &  73.8  & 13.0  \\ %&  73.83  & 13.01& 77.67 &	8.00
    \textbf{\textsc{MFCCs (flat)} } & $10^{-2}\ast$	 & 78.3 & 11.5  \\ %& 78.33 & 11.46& 77.67 &	8.00
    \textbf{\textsc{Spectrogram (flat)} } & $10^{-3}\ast$	 & \textbf{82.3} & 11.4 \\ %\textbf{82.33} & 11.43 & 77.67 &	8.00
    \textbf{\cmp  functionals}  & $10^{-4}\ast$	 &  79.5  & 12.2 \\ %&  79.50  & 12.23 & 80.83	& 11.67
    \textbf{\egm functionals}   & $10^{-3}\ast$	 &  69.0  & 20.1 \\ %&  69.00  & 20.10 & 76.50 &	14.09 
    \textbf{\ds }                & $10^{-6}\ast$  &  66.7  & 10.9  \\ %&  66.67  & 10.94 &  66.67  & 10.94 & -- & --
    \hline 	
     \multicolumn{4}{c}{\bf Features + Neural Networks } \\ \hline  
    \textbf{\textsc{LLDs} + LSTM-RNN} & $10^{-4}\star$	 &  63.3  & 11.0 \\ %&  63.33  & 11.02
    \textbf{\textsc{LLDs} + CRNN} & $10^{-1}\star$	 &  56.3  & 2.6 \\ % &  56.33  & 2.61
    \textbf{\textsc{MFCCs} + LSTM-RNN} & $10^{-4}\star$  &  78.7  & 15.4 \\ %&  78.67  & 15.38
    \textbf{\textsc{MFCCs} + CRNN} & $10^{-5}\star$   &  81.2  & 15.1\\    % &  81.17  & 15.10
    \textbf{\textsc{Spectrogram} + CRNN} & $5^{-3}\star$  &  \textbf{84.0} & 9.9 \\   % \textbf{84.00} & 9.85
    \bottomrule
    \end{tabular}
    }
    \vspace{-2.0em}
\end{table}

For the more complex task of \textbf{stroke success} prediction, we conducted research on the separated data by sex, achieving a UAR of $58.3\,\%$ (LLD + CRNN) and $61.3\,\%$ (MFCC + CRNN), respectively for female and male players. Both together resulted in $60.3\,\%$ (spectrogram + CRNN). 
Overall, spectrograms + CRNN demonstrated consistently strong performance, followed by the other CRNN approaches. 
In contrast, of the expert-knowledge features, only \cmp achieved competitive results, while \egm and \dss again underperformed. The mean and middle feature vectors of any basic feature set in combination with SVM did not result in a UAR higher than by-chance level ($50\,\%$). 
The LDD and spectrogram features concatenated (flattened) along the temporal axis achieved a UAR of $56.2\,\%$ ($\pm 4.2$) and $55.7\,\%$ ($\pm 3.0$). Overall, these results suggest that female and male stroke success are nearly equally possible to predict on acoustic cues, and basic features seem to provide the most predictive representations for this task. Furthermore, relevant information is lost through selection and temporal aggregation, so that adequate models ought to be able to counteract this.
% $58.33\,\%$, $61.33\,\%$, $60.33\.\%$
% $56.17\,\%$ ($\pm 4.17$) and $55.67\,\%$ ($\pm 3.04$)

\begin{table}[t!]
 	\caption{
 	    Results over player-independent 5-fold cross validation on the prediction of score.% on the \score data set.
      	$C$: Complexity parameter of the SVM/SVR, displayed from $10^{-5}$ to $10^{-3}$.
      	$HP$: Set of CRNN hyperparameter from a moderate parameter search over all combinations: 
      	$I$: batch size of 16 and a learning rate of 0.00005;
        $II$: batch size of 16 and a learning rate of 0.0001;
        $III$: batch size of 16 and a learning rate of 0.001;
        $IV$: batch size of 32 and a learning rate of 0.001;
        $V$: batch size of 64 and a learning rate of 0.0001;
        $VI$: batch size of 64 and a learning rate of 0.00001.
		We report the averaged ($\O$) \textbf{UAR}: Unweighted Average Recall and the standard deviation ($\pm$). 
		}
    \vspace{-0.8em}
	\label{tab:results_score}
\centering
  \resizebox{\columnwidth}{!}{
    \begin{tabular}{c|cccccc}
    \toprule
    & \multicolumn{2}{c}{\textbf{women}} & \multicolumn{2}{c}{\textbf{men}} & \multicolumn{2}{c}{\textbf{combined}} \\
     & \O [\%] & $\pm$  [\%] & \O [\%] & $\pm$ [\%] & \O [\%] & $\pm$ [\%] \\   
    \hline
    \hline
     $C$    & \multicolumn{6}{c}{\bf  \cmp  functionals + SVM} \\ \hline                         
    $10^{-5}$&  \textbf{58.3}    &     4.4     &      55.0  & 5.3  &56.5&  2.2\\
    $10^{-4}$&   57.3    &    3.3      &     54.7  & 4.9  &56.8&  1.4\\
    $10^{-3}$&   55.3     &   2.9       &    57.3  & 8.5  &54.8&  2.9\\
    \cline{2-7}
    & \multicolumn{6}{c}{\bf \egm functionals + SVM}\\
    \cline{2-7}
    $10^{-5}$	&	50.7	&	1.3	&	51.0	&	2.7	&	50.7	&	1.1	\\
    $10^{-4}$	&	52.3	&	4.0	&	51.7	&	3.2	&	50.2	&	1.9	\\
    $10^{-3}$	&	53.0	&	3.7	&	51.7	&	6.8	&	53.2	&	5.4	\\
    \cline{2-7}
    & \multicolumn{6}{c}{\textbf{\ds + SVM}} \\
    \cline{2-7}
    $10^{-5}$	&	49.7	&	5.1	&	55.0	&	5.4	&	51.8	&	4.0	\\
    $10^{-4}$	&	51.0	&	5.4	&	58.3	&	4.2	&	52.2	&	2.5	\\
    $10^{-3}$	&	53.0	&	4.9	&	57.0	&	6.9	&	53.3	&	2.9	\\
    \hline
    \hline
    $HP$             & \multicolumn{6}{c}{\textbf{\textsc{Lld} + CRNN}}\\
    \hline
    $I$     &   56.3   &	2.7    & 56.3 &	6.4    & 55.8	&   4.5  \\   %best all & man   
    $II$	&   \textbf{58.3}   &	3.7	& 54.0	&   2.7	& 55.2	&   3.4	\\ % best woman
    $IV$	&   58.0   &	1.6	& 53.7	&   2.2	& 53.3	&   3.0	\\    % for no reason
    \cline{2-7}
    & \multicolumn{6}{c}{\textbf{\textsc{Mfcc} + CRNN}}\\
    \cline{2-7}

    $I$	&   52.3	&   2.5	&   53.7	&   5.3    &	58.7	&   2.3	\\
    $II$    &   51.7	&   2.4    &   56.7	&   7.2    &   55.5	&   3.2 \\       
    $IV$	&	54.3	&   2.5   	&   \textbf{61.3}	&   6.5    &	57.2	&   4.1	\\  
    \cline{2-7}
    & \multicolumn{6}{c}{\textbf{\textsc{Spectrograms} + CRNN}}\\
    \cline{2-7}
    $III$    & 57.7	&   2.0 &   61.0 &   4.0  &   57.8	& 5.1 \\       
    $V$	    & 55.7	&   5.6	&   59.3 &	 7.0	&	\textbf{60.3}	& 3.8	\\
    $VI$	& 56.3	&   3.9	&	60.0 &   5.4	&	57.2	& 5.3	\\ 
    \bottomrule
    \end{tabular}
    }
    \vspace{-2em}
\end{table}

%\section{Limitation and Future Work\label{sectionLimitations}}
% 0) harder to model immediate effect then the final match result 1) A larger data basis means more variety in grunts through the shire amount and the diversity in players that could be added. 2)Reducing the disturbing noises can be fixed partially through adjusting the duration of the audio files exactly to the grunts' lengths. 3)0 Disturbing noises, such as racket hits, appear in parallel to the grunt. Counter by audio denoising, 4) Not only does the volume and frequency of grunts differ with the progress of the game, they also differ with the situations the players are faced with.     Serve, forehand and backhand grunts' fundamental frequency differs significantly~\cite{raine2017tennis}. Introduce more labels to get these situations categorized individually.

\vspace{-1.em}
\section{Conclusion\label{sec:Conclusion}}
\vspace{-0.2em}
In this paper, we introduced \score -- the first audio-visual database suitable for automatic analysis of non-verbal tennis grunts. It comprises 600 female and male grunts of professional tennis player in real tennis matches and two types of balanced labels. Further, we conducted experiments to predict both, utilising a wide range of well-established low-level, expert, and deep audio representations, and machine learning methods. Our models predict sex well, with an average UAR of $84.0\,\%$. Separated by sex, the models predicting if an immediate point was scored after a stroke achieved a UAR of $60.3\,\%$, indicating that tennis grunts contain acoustic patterns related to physiological characteristics of the player. In future work, we plan to investigate whether audio denoising and an even larger amount of data will increase the prediction quality. Furthermore, we want to study different variations of grunting as an acoustic, non-verbal communication between players; for example, to find out if and how much it affects the self-confidence of the opponent.

\vspace{-0.5em}
\section{Acknowledgement}
\vspace{-0.5em}
This project has received funding from KIrun -- AI in sport using audio analysis (German BMWi by ZIM grant No.16KN069402).
%German BMWi by ZIM grant No.16KN069402  (KIrun)

%\vspace{-0.5em}

% BMW Group research. 
\bibliographystyle{IEEEtran}
\vspace{-0.6em}
\bibliography{mybib}
\vspace{-0.6em}
\end{document}